\documentclass[twoside]{article}
\usepackage{fleqn,espcrc2}

\usepackage{graphicx}
\usepackage{rotating}

\def\ga{\mathrel{\raise.3ex\hbox{$>$\kern-.75em\lower1ex\hbox{$\sim$}}}}
\def\la{\mathrel{\raise.3ex\hbox{$<$\kern-.75em\lower1ex\hbox{$\sim$}}}}
\def\he#1{\hbox{${}^{#1}$He}}
\def\li#1{\hbox{${}^{#1}$Li}}

\def\beq{\begin{equation}}
\def\eeq{\end{equation}}

\newcommand{\AmS}{{\protect\the\textfont2
  A\kern-.1667em\lower.5ex\hbox{M}\kern-.125emS}}

\title{Big Bang Nucleosynthesis \vskip -.5in
{\small \rightline{UMN-TH-1753/99}
\rightline{TPI-MINN-99/15}
\rightline{astro-ph/9903309}
\rightline{March 1999}}
\vskip 0in
}

\author{Keith A. Olive\address{Theoretical Physics Institute, School of Physics and Astronomy, 
University of Minnesota,
    Minneapolis MN 55455, USA}%
   \address{Summary of talk delivered at the 19th Texas Symposium on
Relativistic Astrophysics and Cosmology, Paris, France, December 1998.}    
}

\begin{document}

\begin{abstract}
A brief review of standard big bang nucleosynthesis theory and
the related observations of the light element isotopes is presented.
Implications of BBN on chemical evolution and constraints on
particle properties will also be discussed.
\vspace{1pc}
\end{abstract}

\maketitle

\section{INTRODUCTION}

The standard model \cite{wssok,sark1,oc2} of big bang nucleosynthesis (BBN)
is based on the inclusion of an extended nuclear network into a homogeneous
and isotropic FRLW cosmology.  There is now sufficient data to define the
standard model in terms of a single parameter, namely the baryon-to-photon
ratio, $\eta$. Other factors, such as the uncertainties in reaction rates,
and the neutron mean-life can be treated by standard statistical and
Monte Carlo techniques \cite{kr,hata1,sark2} to make predictions (with
specified uncertainties) of the abundances of the light elements, D, \he3,
\he4, and
\li7. Even the number of neutrino flavors, $N_\nu$, which has long
been treated as a parameter can simply be set (=3) in defining the
standard model.  

In this review, I will compare the predictions of BBN with the available
observational determinations of the light element abundances and test for
concordance. I will also discuss the implications of these results on 
the Galactic chemical evolution of the light elements and on limits to
particle properties.

\subsection{Historical Aside}

It is important to bear in mind that there has always been an intimate
connection between BBN and the microwave background as a key test to the
standard big bang model. Indeed, it was the formulation of BBN which
predicted the existence of the microwave background radiation
\cite{gamo}.  The argument is rather simple. BBN requires temperatures
greater than 100 keV, which according to the standard model
time-temperature relation, $t_{\rm s} T^2_{\rm MeV} = 2.4/\sqrt{N}$, where
$N$ is the number of relativistic degrees of freedom at temperature
$T$, corresponds to timescales less than about 200 s. The typical
cross section for the first link in the nucleosynthetic chain is
\beq 
\sigma v (p + n \rightarrow D + \gamma) \simeq 5 \times 10^{-20} 
{\rm cm}^3/{\rm s}
\eeq
This implies that it was necessary to achieve a density
\beq
n \sim {1 \over \sigma v t} \sim 10^{17} {\rm cm}^{-3}
\eeq
The density in baryons today is known approximately from the density of
visible matter to be ${n_B}_o \sim 10^{-7}$ cm$^{-3}$ and since
we know that that the density $n$ scales as $R^{-3} \sim T^3$, 
the temperature today must be
\beq
T_o = ({n_B}_o/n)^{1/3} T_{\rm BBN} \sim 10 {\rm K}
\eeq
thus linking two of the most important tests of the Big Bang theory.

\section{THEORY}

Conditions for the synthesis of the light elements were attained in the
early Universe at temperatures  $T \la $ 1 MeV.  
At somewhat higher temperatures, weak interaction rates were
in equilibrium. In particular, the processes
\begin{eqnarray}
n + e^+ & \leftrightarrow  & p + {\bar \nu_e} \nonumber \\
n + \nu_e & \leftrightarrow  & p + e^- \nonumber \\
n  & \leftrightarrow  & p + e^- + {\bar \nu_e} \nonumber 
\end{eqnarray}
fix the ratio of
number densities of neutrons to protons. At $T \gg 1$ MeV, $(n/p) \simeq
1$.

 As the temperature fell
and approached the point where the weak interaction rates were no longer
fast enough to maintain equilibrium, the neutron to proton ratio was given
approximately by the Boltzmann factor,
$(n/p)
\simeq e^{-\Delta m/T}$, where $\Delta m$ is the neutron-proton mass
difference. The final abundance of \he4 is very sensitive to the $(n/p)$
ratio
\beq
Y_p = {2(n/p) \over \left[ 1 + (n/p) \right]} \approx 0.25
\label{ynp}
\eeq
 Freeze out occurs at slightly less than an MeV resulting in $(n/p) \sim
1/6$ at this time.

The nucleosynthesis chain begins with the formation of Deuterium
through the process, $p+n \rightarrow$ D $+ \gamma$.
However, because the large number of photons relative to nucleons,
$\eta^{-1} = n_\gamma/n_B \sim 10^{10}$, Deuterium production is delayed
past the point where the temperature has fallen below the Deuterium
binding energy, $E_B = 2.2$ MeV (the average photon energy in a blackbody
is ${\bar E}_\gamma \simeq 2.7 T$).  This is because there are many
photons in the exponential tail of the photon energy distribution with
energies $E > E_B$ despite the fact that the temperature or ${\bar
E}_\gamma$ is less than $E_B$. During this delay, the neutron-to-proton
ratio drops to $(n/p) \sim 1/7$.

\begin{figure}[htbp]
\hspace{0.5truecm}
\includegraphics[width=15pc]{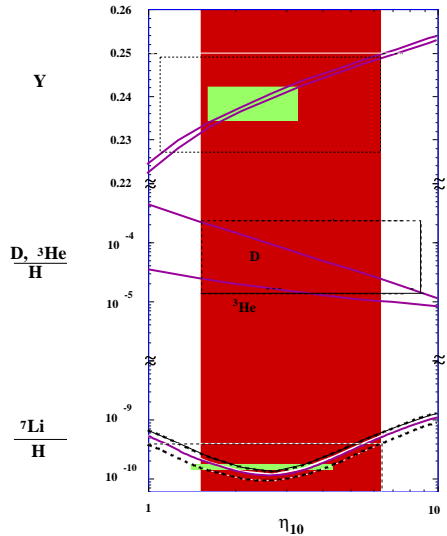}
\caption{{The light element abundances from big bang
nucleosynthesis as a function of $\eta_{10} = 10^{10}\eta$.}}
\label{nuc8}
\end{figure}

The dominant product of big bang nucleosynthesis is \he4 resulting in an
abundance of close to 25\% by mass. Lesser amounts of the other light
elements are produced: D and \he3 at the level of about $10^{-5}$ by
number,  and \li7 at the level of $10^{-10}$ by number. The resulting
abundances of the light elements are shown in Figure \ref{nuc8}, which
concentrate on the range in
$\eta_{10}$ between 1 and 10.  The curves for the \he4 mass fraction,
$Y$, bracket the computed range based on the uncertainty of the neutron
mean-life which  has been taken as  $\tau_n = 887 \pm 2$ s. 
 Uncertainties in the produced \li7 
abundances have been adopted from the results in Hata et al.
\cite{hata1}. Uncertainties in D and
\he3 production are small on the scale of this figure. 
The  boxes correspond
to the observed abundances and will be discussed below. 

\section{Data}

\subsection{\he4}

The primordial \he4 abundance is best
determined from observations  of HeII
$\rightarrow$ HeI recombination lines in extragalactic HII 
(ionized hydrogen) regions.
There is now a good collection of abundance information on the \he4 mass
fraction, $Y$, O/H, and N/H in over 70 
\cite{p,evan,iz} such regions. 
Since \he4 is produced in stars along with heavier elements such as Oxygen,
it is then expected that the primordial abundance of \he4 can be determined
from the intercept of the correlation between $Y$ and O/H, namely $Y_p =
Y({\rm O/H} \to 0)$.   A detailed analysis of the data including that in
\cite{iz} found an intercept corresponding to a primordial abundance  
 $Y_p  = 0.234 \pm 0.002 \pm 0.005$ \cite{ost3}.  This was updated to
include the most recent results of \cite{iz2} in \cite{fdo2}.  The result
(which is used in the discussion below) is
\beq
Y_p = 0.238 \pm 0.002 \pm 0.005
\label{he4}
\eeq
The first uncertainty is purely statistical and the second uncertainty is
an estimate of the systematic uncertainty in the primordial abundance
determination \cite{ost3}.  The solid box for
\he4 in Figure \ref{nuc8} represents the range (at 2$\sigma_{\rm stat}$)
from (\ref{he4}). The dashed box extends this by including the systematic
uncertainty. A
somewhat lower primordial abundance of  $Y_p = 0.235 \pm .003$ is found
by restricting to the 36 most metal poor regions \cite{fdo2}.  
These results are consistent with those of a Bayesian analysis \cite{hos}
based on the 32 points of lowest metallicity. These have been used to
calculate a likelihood function for which the peak occurs at $Y_p
= 0.238$ and the most likely spread is $w = 0.009$. The 95\% CL upper
limit to $Y_p$ in this case is 0.245.  For further details on this approach
see \cite{hos}.

\begin{figure}[htbp]
\hspace{0.5truecm}
\includegraphics[width=15pc]{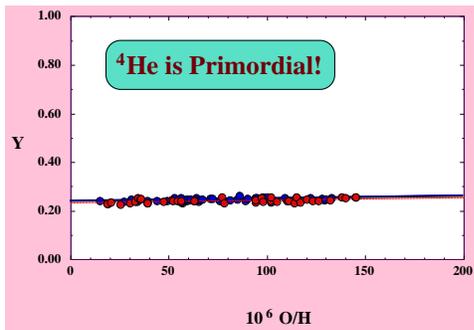}
\caption{{The \he4 mass fraction as determined in extragalactic H II
regions as a function of O/H.}}
\label{he1}
\end{figure}

A global view of the \he4 data is shown in Figure \ref{he1}.
What should be absolutely apparent from this figure is the primordial
nature of \he4.  There are no observations to date which indicate a \he4
abundance which is significantly below 23\% to 24\%.  In particular, even
in systems in which an element such as Oxygen, which traces stellar
activity, is observed at extremely low values (compared with the solar
value of O/H $\approx 8.5 \times10^{-4}$), the \he4 abundance is nearly
constant.  This is far different from all other element abundances (with
the exception of \li7 as we will see below).  For example, in Figure
\ref{no}, the N/H vs. O/H correlation is shown.  As one can clearly see,
the abundance of N/H goes to 0, as O/H goes to 0, indicating a stellar
source for Nitrogen.

\begin{figure}[htbp]
\hspace{0.5truecm}
\includegraphics[width=15pc]{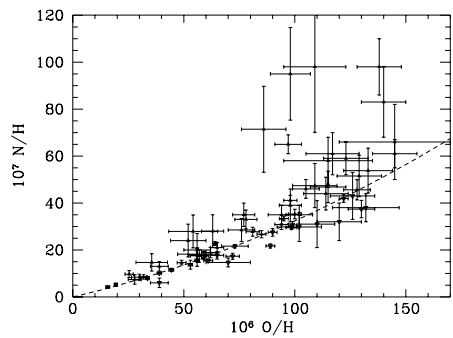}
\caption{{The Nitrogen and Oxygen abundances in the same extragalactic
HII regions with observed \he4 shown in Figure \protect\ref{he1}.}}
\label{no}
\end{figure}

\begin{figure}[htbp]
\hspace{0.5truecm}
\includegraphics[width=15pc]{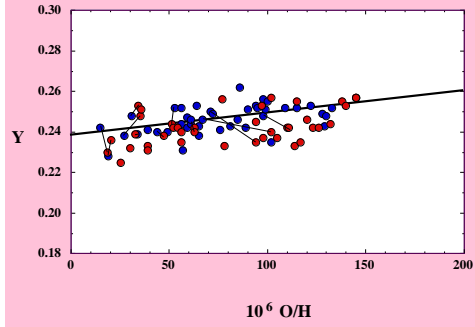}
\caption{{The Helium (Y) and Oxygen (O/H) abundances 
 in extragalactic HII regions, 
from refs. \protect\cite{p,evan}  and from ref.
\protect\cite{iz2}.   Lines connect the  same regions
observed by different groups. The regression shown leads to the primordial
\he4 abundance given in Eq. (\protect\ref{he4}). }}
\label{he2}
\end{figure}

A more useful plot of the \he4 data is shown in Figure \ref{he2}.
Here one sees the correlation of \he4 with O/H and the linear regression
which leads to primordial abundance given in Eq. (\ref{he4}).

\subsection{\li7}

The abundance of \li7 has been determined by observations of over 100
hot, population-II stars, and is found to have a very
nearly  uniform abundance \cite{sp}. For
stars with a surface temperature $T > 5500$~K
and a metallicity less than about
1/20th solar (so that effects such as stellar convection may not be
important), the  abundances show little or no dispersion beyond that which
is consistent with the errors of individual measurements.
Indeed, as detailed in ref. \cite{mol,rnb}, much of the work concerning
\li7 has to do with the presence or absence of dispersion and whether
or not there is in fact some tiny slope to a [Li] = $\log$ \li7/H + 12 vs.
T or [Li] vs. [Fe/H]  relationship ([Fe/H] is the log of the Fe/H ratio
relative to the solar value).

\begin{figure}[htbp]
\hspace{0.5truecm}
\includegraphics[width=15pc]{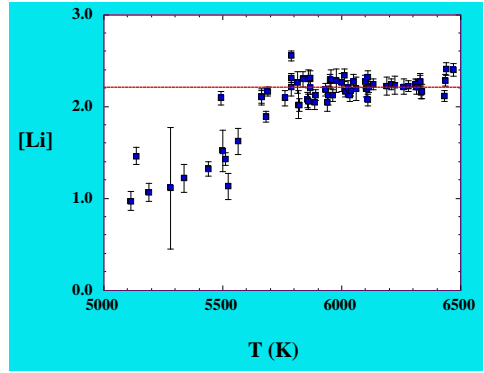}
\caption{{The Li abundance in halo stars with [Fe/H] $<$ -1.3,
as a function of surface temperature. The dashed line shows the value of
the weighted mean of the plateau data.}}
\label{fig:lit}
\end{figure}

 When the Li data from stars with [Fe/H] $<$ 
-1.3 is plotted as a function of surface temperature, one sees a 
plateau emerging for $T > 5500$ K as shown in Figure \ref{fig:lit} for 
the data taken from ref. \cite{mol}.
As one can see from the figure, at high temperatures, where the convection
zone does not go deep below the surface, the Li abundance is uniform.
At lower surface temperatures, the surface abundance of Li is depleted as
Li is dragged through the hotter interior of the star and destroyed.
The lack of dispersion in the plateau region is evidence that this abundance
is indeed primordial (or at least very close to it).  

I will use the value given in ref. \cite{mol} 
as the best estimate
for the mean \li7 abundance and its statistical uncertainty in halo stars 
\beq
{\rm Li/H = (1.6 \pm 0.1 ) \times 10^{-10}}
\label{li}
\eeq
The small error is is statistical and is due to the large number of stars
in which \li7 has been observed. The solid box for \li7 in Figure
\ref{nuc8} represents the 2$\sigma_{\rm stat}$ range from (\ref{li}).
There is however an important source of systematic error due to the
possibility that Li has been depleted in these stars, though the lack of
dispersion in the Li data limits the amount of depletion.  In addition, 
standard stellar models\cite{del} predict that any depletion of \li7 would
be  accompanied by a very severe depletion of \li6.  Until recently, 
\li6 had never been observed in
hot pop II stars. The observation\cite{li6o} of 
\li6 (which turns out to be
consistent with its origin in cosmic-ray nucleosynthesis) 
is  another good indication that
\li7 has not been destroyed in these stars \cite{li6,pin,vcof}.

Aside from the big bang, Li is  produced together with Be and B in
accelerated particle interactions such as cosmic ray spallation of C,N,O by
protons and $\alpha$-particles.  Li is also produced by  $\alpha-\alpha$
fusion.  Be and B have been observed in these same pop II stars and in
particular there are a dozen or so stars in which both Be and \li7 have
been observed.  Thus Be (and B though there is still a paucity of
data) can be used as a consistency check on primordial Li \cite{fossw}. 
Based on the Be abundance found in these stars, 
one can conclude that no more than 10-20\% of 
the \li7 is due to cosmic ray nucleosynthesis leaving the remainder
(an abundance near $10^{-10}$) as primordial.
A similar conclusion was reached in \cite{rnb}.
The dashed box in Figure \ref{nuc8}, accounts for
 the possibility that as much as half
of the primordial \li7 has been
destroyed in stars, and that as much as 20\% of the observed \li7   may
have been produced in cosmic ray collisions rather than in the Big Bang.
 For \li7, the uncertainties are clearly dominated by
systematic effects.

\section{Likelihood Analyses}

At this point, having established the primordial abundance of at least two
of the light elements, \he4 and \li7, with reasonable certainty, it is
possible to test the concordance of BBN theory with observations.
Monte Carlo techniques have proven to be a useful form of analysis in this
regard
\cite{kr,hata1,sark2}.  Two
elements are sufficient for not only constraining the one parameter 
theory of BBN, but also for testing for consistency \cite{fo}. 
The procedure begins by establishing likelihood functions for the theory and
observations. For example, for \he4, the theoretical likelihood 
function takes the
form
\beq
L_{\rm BBN}(Y,Y_{\rm BBN}) 
  = e^{-\left(Y-Y_{\rm BBN}\left(\eta\right)\right)^2/2\sigma_1^2}
\label{gau}
\eeq
where $Y_{\rm BBN}(\eta)$ is the central value for the \he4 mass fraction
produced in the big bang as predicted by the theory at a given value of $\eta$.
$\sigma_1$ is the uncertainty in that  value derived from the Monte Carlo
calculations \cite{hata1} and is a measure of the theoretical 
uncertainty in the
big bang calculation. Similarly one can write down an expression for the
observational likelihood function. Assuming Gaussian errors,
the likelihood function for the observations would
take a form similar to that in (\ref{gau}).

\begin{figure}[htbp]
\hspace{0.5truecm}
\includegraphics[width=15pc]{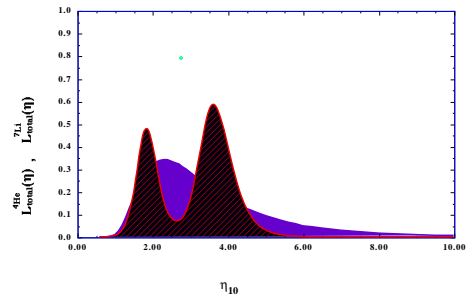}
\caption{{Likelihood distribution for each of \he4 and
\li7, shown as a  function of $\eta$.  The one-peak structure of the \he4
curve corresponds to the monotonic increase of $Y_p$ with $\eta$, while
the two peaks for \li7 arise from the minimum in the  \li7 abundance
prediction.}}
\label{fig:fig1}
\end{figure}

A total likelihood 
function for each value of $\eta$ is derived by
convolving the theoretical
and observational distributions, which for \he4 is given by
\begin{eqnarray}
& {L^{^4{\rm He}}}_{\rm total}(\eta) &  \nonumber \\
 & = \int dY L_{\rm BBN}\left(Y,Y_{\rm BBN}\left(\eta\right)\right) 
L_{\rm O}(Y,Y_{\rm O}) &
\label{conv}
\end{eqnarray}
An analogous calculation is performed \cite{fo} for \li7. 
The resulting likelihood
functions from the observed abundances given in Eqs. (\ref{he4}) 
  and (\ref{li})
is shown in Figure \ref{fig:fig1}. As one can see 
there is very good agreement between \he4 and \li7 in the range
of $\eta_{10} \simeq$ 1.5 -- 5.0. The double peaked nature of the \li7
likelihood function is due to the presence of a minimum in the
 predicted lithium abundance.  For a given observed value of \li7, there
are two likely values of $\eta$.

\begin{figure}[htbp]
\hspace{0.5truecm}
\includegraphics[width=15pc]{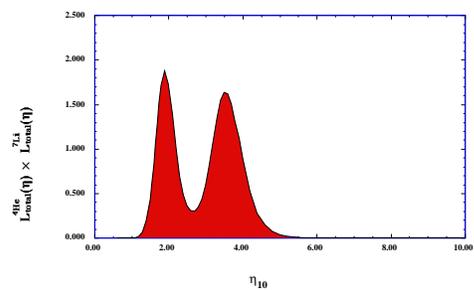}
\caption{{Combined likelihood for simultaneously fitting \he4 and \li7,
as a function of $\eta$.}}
\label{fig:fig2}
\end{figure}

The combined likelihood, for fitting both elements simultaneously,
is given by the product of the two functions in Figure \ref{fig:fig1}
and is shown in Figure \ref{fig:fig2}. The 95\% CL region covers the
range $1.55 < \eta_{10} < 4.45$, with the two peaks occurring at
$\eta_{10} = 1.9$ and 3.5. This range corresponds to values of
$\Omega_B$ between
\beq
0.006 < \Omega_B h^2 < .016
\label{omega}
\eeq 

\section{More Data}

\subsection{D and \he3}

Because there are no known astrophysical sites for the production of
Deuterium, all observed D is assumed to be primordial. As a result,
any firm determination of a Deuterium abundance establishes an upper bound
on $\eta$ which is robust.

Deuterium abundance information is available from several astrophysical
environments, each corresponding to a different evolutionary epoch. In the
ISM, corresponding to the present epoch, an often quoted measurement for
D/H is \cite{linetal}
\beq
{\rm (D/H)_{ISM}} = 1.60\pm0.09{}^{+0.05}_{-0.10} \times 10^{-5}
\eeq
This measurement allow us to set the upper limit to $\eta_{10} < 9$
and is shown by the lower right of the solid box
in Figure \ref{nuc8}. There are however, serious questions regarding
homogeneity of this value in the ISM. There may be evidence for
considerable dispersion in D/H \cite{disp,vm} as is the case with \he3
\cite{bbbrw}.

The solar abundance of D/H is inferred from two
distinct measurements of \he3. The solar wind measurements of \he3 as well
as the low temperature components of step-wise heating measurements of
\he3 in meteorites yield the presolar (D + \he3)/H ratio, as D was 
efficiently burned to \he3 in the Sun's pre-main-sequence phase.  These
measurements  indicate that \cite{scostv,geiss}
$
{\rm \left({(D +~^3He) / H} \right)_\odot = (4.1 \pm 0.6 \pm 1.4) \times
10^{-5}}
$.
 The high temperature components in meteorites are believed to yield the
true solar \he3/H ratio of \cite{scostv,geiss}
$
{\rm \left({~^3He / H} \right)_\odot = (1.5 \pm 0.2 \pm 0.3) \times
10^{-5}}
$.
The difference between these two abundances reveals the presolar D/H ratio,
giving,
\beq
{\rm (D/H)_{\odot}} \approx (2.6 \pm 0.6 \pm 1.4) \times 10^{-5}
\eeq
This value for presolar D/H is consistent with measurements of surface
abundances of HD on Jupiter
 D/H = $2.7 \pm 0.7 \times 10^{-5}$ 
\cite{nie}. 

Finally, there have been several reported measurements of 
D/H in high redshift quasar absorption systems. Such measurements are in
principle capable of determining the primordial value for D/H and hence
$\eta$, because of the strong and monotonic dependence of D/H on $\eta$.
However, at present, detections of D/H  using quasar absorption systems
do not yield a conclusive value for D/H.  As such, it should be cautioned 
that these values may not
turn  out to represent the true primordial value and it is very unlikely 
that both are primordial and indicate an inhomogeneity \cite{cos2}
(a large scale inhomogeneity of the magnitude required to placate all
observations is excluded by the isotropy of the microwave background
radiation). The first of these measurements
\cite{quas1} indicated a rather high D/H ratio, D/H $\approx$ 1.9 -- 2.5
$\times 10^{-4}$.  Other  high D/H ratios were reported in \cite{quas3}. 
More  recently, a similarly high value of D/H = 2.0 $\pm 0.5 \times
10^{-4}$ was reported in a relatively low redshift system (making it less
suspect to interloper problems) \cite{webb}. This was confirmed in
\cite{thigh} where a 95\% CL lower bound to D/H was reported as 8 $\times
10^{-5}$.
 However, there are reported low values of D/H in other such systems
\cite{quas2} with values of D/H originally reported as low as
$\simeq 2.5
\times 10^{-5}$, significantly lower than the ones quoted above. 
The abundance in these systems has been revised upwards to about 3.4 $\pm
0.3 \times 10^{-5}$ \cite{bty}.
However, it was also noted \cite{lev} that when using mesoturbulent
models to account for the velocity field structure in these systems, the
abundance may be somewhat higher (3.5 -- 5 $\times 10^{-5}$). This may be
quite significant, since at the upper end of this range (5 $\times
10^{-5}$) all of the element abundances are consistent as will be
discussed shortly. I will not enter into the debate as to which if any of
these observations may be a better representation of the true primordial
D/H ratio.  The status of these observations are more fully reviewed in
\cite{vm}. The upper range of quasar absorber D/H is shown by the dashed
box in Figure
\ref{nuc8}.

There are also several types of \he3 measurements. As noted above, meteoritic
extractions yield a presolar value for \he3/H.
In addition, there are several ISM measurements of \he3 in galactic HII
regions \cite{bbbrw} which show a wide dispersion which may be indicative 
of pollution or a bias \cite{orstv}
$
 {\rm \left({^3He/ H} \right)_{HII}} \simeq 1 - 5 \times 10^{-5}
$.
There is also a recent ISM measurement of \he3 \cite{gg}
with
$
 {\rm \left({^3He /H} \right)_{ISM}} = 2.1^{+.9}_{-.8} \times 10^{-5}
$.
  Finally there are observations of \he3 in planetary
nebulae \cite{rood} which show a very high \he3 abundance of 
\he3/H $\sim 10^{-3}$.  None of these determinations represent the
primordial \he3 abundance, and as will be discussed below, their relation to
the primordial abundance is heavily dependent on both stellar and chemical
evolution.

\section{More Analysis}

 It is interesting to
compare the results from the likelihood functions of \he4 and \li7 with
that of D/H.  
Since  D and \he3 are monotonic functions of $\eta$, a prediction for 
$\eta$, based on \he4 and \li7, can be turned into a prediction for
D and \he3.  
 The corresponding 95\% CL ranges are D/H  $= (4.1 - 25)  \times
10^{-5}$ and \he3/H $= (1.2 - 2.6)  \times 10^{-5}$.
If we did have full confidence in the measured value of D/H in 
quasar absorption
systems, then we could perform the same statistical analysis 
using \he4, \li7, and
D. To include D/H, one would
proceed in much the same way as with the other two light elements.  We
compute likelihood functions for the BBN predictions as in
Eq. (\ref{gau}) and the likelihood function for the observations.
These are then convolved as in Eq.  (\ref{conv}). 

\begin{figure}[htbp]
\hspace{0.5truecm}
\includegraphics[width=15pc]{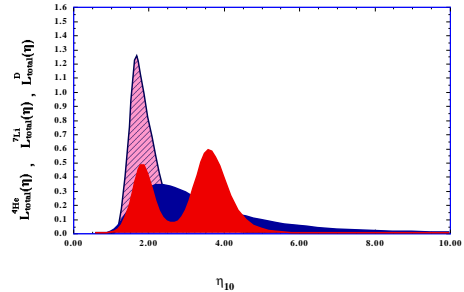}
\caption{As in Figure \protect\ref{fig:fig1}, with the addition of the
likelihood  distribution for D/H assuming ``high" D/H.}
\label{D1}
\end{figure}

\begin{figure}[htbp]
\hspace{0.5truecm}
\includegraphics[width=15pc]{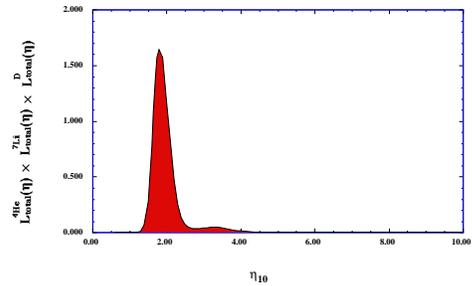}
\caption{Combined likelihood for simultaneously fitting 
\he4 and \li7, and D 
as a function of $\eta$ from Figure \protect\ref{D1}.}
\label{D2}
\end{figure}

 Using D/H = $(2.0 \pm 0.5) \times 10^{-4}$ as indicated in the high
D/H systems, we can plot the three likelihood functions including
$L^{{\rm D}}_{\rm total}(\eta)$  in Figure \ref{D1}.
  It is indeed startling how the three peaks, for
D, \he4 and \li7 are in excellent agreement with each other.  In Figure
\ref{D2},  the combined distribution is shown.
We now  have a very clean distribution and prediction 
for $\eta$, $\eta_{10}  = 1.8^{+1.6}_{-.3}$ corresponding to $\Omega_B h^2
= .007^{+.005}_{-.001}$.  
The absence of any overlap with the high-$\eta$ peak of the \li7
distribution has considerably lowered the upper limit to $\eta$. 
Overall, the concordance limits in this case are dominated by the 
Deuterium likelihood function.


\begin{figure}[htbp]
\hspace{0.5truecm}
\includegraphics[width=15pc]{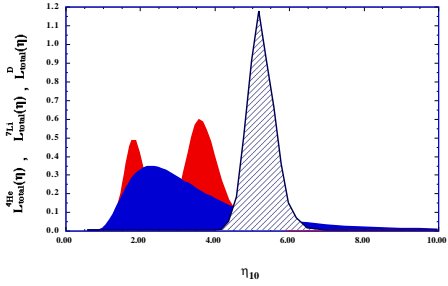}
\caption{As in Figure \protect\ref{D1}, with the likelihood 
distribution for low D/H.}
\label{Dt1}
\end{figure}

\begin{figure}[htbp]
\hspace{0.5truecm}
\includegraphics[width=15pc]{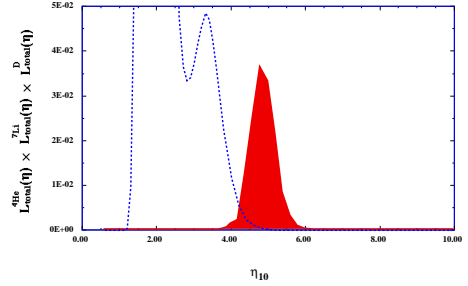}
\caption{Combined likelihood for simultaneously fitting 
\he4 and \li7, and  low D/H
as a function of $\eta$. The dashed curve represents the combined
distribution shown in Figure \protect\ref{D2}.}
\label{Dt2}
\end{figure}

 If instead, we assume that the low value \cite{bty}
of D/H = $(3.4 \pm 0.3) \times 10^{-5}$ is the primordial abundance,
then we can again compare the likelihood distributions as in Figure
\ref{D1}, now substituting the low D/H value. As one can see from Figure
\ref{Dt1}, there  is now hardly any overlap between the D and the \li7
and \he4 distributions.  The combined distribution shown in Figure \ref{Dt2} is
compared with that in Figure \ref{D2}. 
In this case, D/H is just compatible (at the 2 $\sigma$ level) with
the other light elements, and the peak of the likelihood function
occurs at  $\eta_{10}  = 4.8^{+0.5}_{-0.6}$. Though one can not use this
likelihood analysis to prove the correctness of the high D/H measurements
or the incorrectness of the low D/H measurements, the analysis clearly
shows the difference in compatibility between the two values of D/H and
the observational determinations of \he4 and \li7. To {\em make} the low
D/H measurement compatible, one would have to argue for a shift upwards in
\he4 to a primordial value of 0.247 (a shift by 0.009) which is not
warranted at this time by the data, and a \li7 depletion factor of 
about 2, which is close to recent upper limits to the amount of depletion
\cite{cv,pin}.


It is important to recall however, that the true uncertainty in the low
D/H systems might be somewhat larger.  If we allow D/H to be as large as
$5 \times 10^{-5}$, the peak of the D/H likelihood function shifts down to 
$\eta_{10} \simeq 4$.  In this case, there would be a near perfect overlap
with the high $\eta$ \li7 peak and since the \he4 distribution function is
very broad, this would be a highly compatible solution.

\section{Chemical Evolution}

Because we can not directly measure the primordial abundances of any of
the light element isotopes, we are required to make some assumptions
concerning the evolution of these isotopes. As has been discussed above, 
\he4 is produced in stars along with Oxygen and Nitrogen.
\li7 can be destroyed in stars and produced in several
(though still uncertain) environments. D is totally destroyed in the star 
formation process and \he3 is both produced and destroyed in stars with
fairly uncertain yields. It is therefore preferable, if possible
to observe the light element isotopes in a low metallicity 
environment. Such is the case with \he4 and \li7.  These elements are
observed in environments which are as low as 1/50th and 1/1000th solar
metallicity respectively and we can be fairly assured that the abundance
determinations of these isotopes are close to  primordial.  If the quasar
absorption system measurements of D/H stabilize, then this too may be very
close to a primordial measurement.  Otherwise, to match the solar and
present abundances of D and \he3 to their  primordial values requires a
model of galactic chemical evolution.

 The main inputs to chemical evolution models are:
 1) The initial mass function, $\phi(m)$, indicating the 
distribution of stellar masses. Typically, a simple
power law form for the IMF is chosen, $\phi(m) \sim m^{-x}$,
with $x \simeq -2.7$.  This is a fairly good representation of the
observed distribution, particularly at larger masses.
 2) The star formation rate, $\psi$. Typical choices for a SFR
are $\psi(t) \propto \sigma$ or $\sigma^2$ or even a straight exponential
$e^{-t/\tau}$.  $\sigma$ is the fraction of mass in gas, 
$M_{\rm gas}/M_{\rm tot}$. 3) The presence
of infalling or outflowing gas; and of course 4) the stellar yields.  It is 
the latter, particularly in the case of \he3, that is the cause for
so much uncertainty. Chemical evolution models simply set up a series of 
evolution equations which trace desired quantities. 

Deuterium is always a monotonically
decreasing function of time in chemical evolution models.  The degree to
which  D is destroyed, is however a model dependent
question which depends sensitively on the IMF and SFR.
The evolution of \he3 is however considerably more complicated.
Stellar models predict that substantial amounts of \he3 are
produced in stars between 1 and 3 M$_\odot$.  For example, in the
models of Iben and Truran \cite{it} a 1 M$_\odot$ star will yield a 
\he3 abundance which is nearly three times as large as its initial 
(D+\he3) abundance.
It should be emphasized that this prediction is in
fact consistent with the observation of high \he3/H in planetary nebulae
\cite{rood}.

\begin{figure}[htbp]
\hspace{0.5truecm}
\includegraphics[width=15pc]{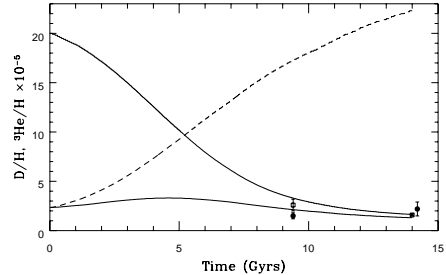}
\vskip -1.6in
\caption{The evolution of D/H and 
\he3/H with time in units of $10^{-5}$. The assumed
primordial abundance of D/H is 2 $\times 10^{-4}$.
The dashed \he3 curve shows the evolution with standard \he3 yields, while
the solid curve shows the effect of the reduced yields of \cite{bm}.}
\label{evol1}
\end{figure}

However, the implementation of standard model \he3 yields 
in chemical evolution models leads to an overproduction of \he3/H
particularly at the solar epoch \cite{orstv,galli}.
While the overproduction is problematic for any initial value of D/H, it
is particularly bad in models with a high primordial D/H.  In Scully
et al. \cite{scov}, a dynamically generated supernovae wind model was
coupled to models of galactic chemical evolution with the aim of reducing
a primordial D/H abundance of 2 $\times 10^{-4}$ to the present ISM value
without overproducing heavy elements and  remaining consistent with the
other observational constraints typically  imposed on such models.
In Figure \ref{evol1}, the evolution of D/H and 
\he3/H is
shown as a function of time in one of the models 
with significant Deuterium destruction factors (see ref \cite{scov} for
details). While such a model can successfully account for the evolution of
D/H (and other standard chemical evolution tracers), as one can
plainly see,
\he3 is grossly overproduced (the Deuterium data is represented by squares
and
\he3 by circles). 

The overproduction of \he3 relative to the solar meteoritic value seems to be a
generic feature of chemical evolution models when \he3 production in low mass
stars is included. This result appears to be independent of the 
chemical evolution
model and is directly related to the assumed stellar yields of \he3.
It has recently been suggested that at least some low mass
stars may indeed be net destroyers of \he3 if one includes
the effects of extra mixing below the conventional convection zone
in low mass stars on the red giant branch \cite{char,bm}. The extra  
mixing does not take place for stars which do not undergo a Helium core
flash (i.e. stars $>$ 1.7 - 2 M$_\odot$ ).  Thus stars with masses {\it
less than} 1.7 M$_\odot$ are responsible for the \he3 destruction. 
Using the yields of Boothroyd and Malaney \cite{bm}, it was shown
\cite{osst} that these reduced \he3 yields in low mass stars can account
for the relatively low solar and present day \he3/H abundances observed.
In fact, in some cases, \he3 was underproduced.  To account for the \he3
evolution and the fact that some low mass stars must be producers 
of \he3 as indicated by the
planetary nebulae data, it was suggested that the new yields apply
only to a fraction (albeit large) of low mass stars \cite{osst,gal}. 
The corresponding evolution \cite{osst} of 
D/H and \he3/H using the reduced yields is shown in Figure \ref{evol1}.

The models of chemical evolution discussed above indicate that it is
possible to destroy significant amounts of Deuterium and remain
consistent with chemical evolutionary constraints.  To do so however, comes
with a price. Large Deuterium destruction factors require substantial
amounts of stellar processing, which at the same time produce heavy
elements.  To keep the heavy element abundances in the Galaxy in check,
significant Galactic winds enriched in heavy elements must be
incorporated.  In fact there is some evidence that enriched winds were
operative in the early Galaxy.  In the X-ray cluster satellites observed
by Mushotzky et al. \cite{mush1} and  Loewenstein and Mushotzky
\cite{mush2} the mean Oxygen abundance was found to be roughly half solar.
This corresponds to a near solar abundance of heavy elements in the
inter-Galactic medium, where apparently little or no star formation has
taken place. 

\begin{figure}[htbp]
\hspace{0.5truecm}
\includegraphics[width=15pc]{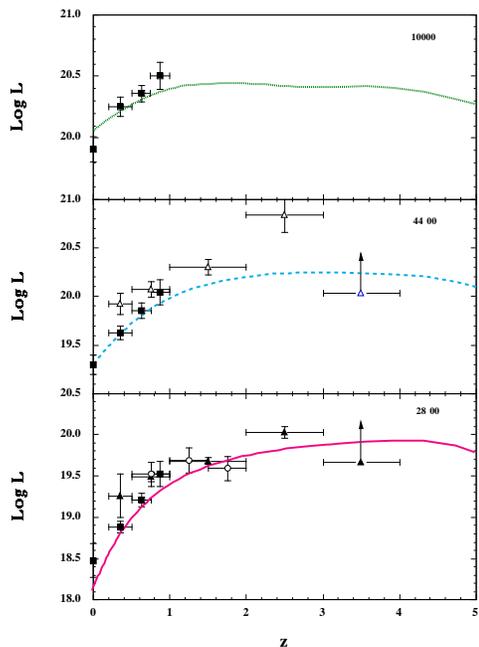}
\caption{The tricolor luminosity  densities (UV, B and IR) at
$\lambda = 0.28, 0.44$ and 1.0 $\mu$m, in units of (h/.5)
WHz$^{-1}$Mpc$^{-3}$ as a function of redshift for a model shown in
\protect\ref{evol1} which destroys significant amounts of D/H. The data
are taken from
\protect\cite{cce}.}
\label{fig:cova}
\end{figure}

If our Galaxy is typical in the Universe, then the models of the type
discussed above would indicate that the luminosity density of the
Universe at high redshift should also be substantial augmented relative
to the present. Recent observations of the luminosity density at high
redshift \cite{cce}  are making it possible for the first time
to test models of cosmic chemical evolution. 
The high redshift observations, are very discriminatory with
respect to a given SFR \cite{cova}.  Models in which the star formation
rate is proportional to the gas mass fraction (these are common place in
Galactic chemical evolution) have difficulties to fit the multi-color
data from $z = 0$ to 1.  This includes many of the successful Galactic
infall models. In contrast, 
models with  a steeply decreasing SFR  are
favored.  In Figure \ref{fig:cova}, the predicted luminosity
density based on the model with evolution shown in Figure \ref{evol1} from
\cite{scov}, as compared with the observations (see ref.
\cite{cova} for details).

While it would be premature to conclude that all models with large
Deuterium destruction factors are favored, it does seem that models which
do fit the high redshift data destroy significant amounts of D/H.  On the
other hand, we can not exclude models which destroy only a small amount
of D/H as Galactic models of chemical evolution. In this case, however the
evolution of our Galaxy is anomalous with respect to the cosmic average.
If the low D/H measurements \cite{quas2,bty}  hold up, then it would
seem that our Galaxy also has an anomalously high  D/H abundance.  That is
we would predict in this case that the present cosmic abundance of D/H is
significantly lower than the observed ISM value.
(This conclusion assumes that the ISM D/H abundance is representative of
the present epoch.  If new results \cite{disp} which show values as low as
a few $\times 10^{-6}$ are more typical, then low primordial D/H would
fit the high redshift data and models with a high degree of D
destruction much better.)   If the high D/H observations 
\cite{quas1,quas3,webb} hold up, we could conclude that our Galaxy is
indeed representative of the cosmic star formation history.

\section{Constraints from BBN}

Limits on particle physics beyond the standard model are mostly sensitive to
the bounds imposed on the \he4 abundance. 
As is well known, the $^4$He abundance
is predominantly determined by the neutron-to-proton ratio just prior to
nucleosynthesis and is easily estimated assuming that all neutrons are
incorporated into \he4.
As discussed earlier, the neutron-to-proton
ratio is fixed by its equilibrium value at the freeze-out of 
the weak interaction rates at a temperature $T_f \sim 1$ MeV modulo the
occasional free neutron decay.  Furthermore, freeze-out is determined by
the competition between the weak interaction rates and the expansion rate
of the Universe
\begin{equation}
{G_F}^2 {T_f}^5 \sim \Gamma_{\rm wk}(T_f) = H(T_f) \sim \sqrt{G_N N} {T_f}^2
\label{comp} \label{freeze}
\end{equation}
where $N$ counts the total (equivalent) number of relativistic particle
species. The presence
of additional neutrino flavors (or any other relativistic species) at 
the time of nucleosynthesis increases the overall energy density
of the Universe and hence the expansion rate leading to a larger 
value of $T_f$, $(n/p)$, and ultimately $Y_p$.  Because of the
form of Eq. (\ref{comp}) it is clear that just as one can place limits
\cite{ssg} on $N$, any changes in the weak or gravitational coupling constants
can be similarly constrained (for a recent discussion see ref. \cite{co}).

As discussed above, the limit on $N_\nu$ comes about via the 
change in the expansion rate given by the Hubble parameter,
\beq
H^2 = {8 \pi G \over 3} \rho = {8 \pi^3  G \over 90} [N_{\rm SM} 
+ {7 \over 8} \Delta N_\nu] T^4
\eeq
when compared to the weak interaction rates. Here $N_{\rm SM}$
refers to the standard model value for N. At $T \sim 1$ MeV,
$N_{\rm SM} = 43/4$. Additional degrees of freedom will 
lead to an increase in the freeze-out temperature eventually leading to
a higher \he4 abundance. In fact, one 
can parameterize the dependence of $Y$ on $N_\nu$ by 
\beq
Y = 0.2262 + 0.0131 (N_\nu - 3) + 0.0135 \ln \eta_{10} 
\label{YY}
\eeq
in the vicinity of $\eta_{10} \sim 2$.  Eq. (\ref{YY}) also shows
the weak (log) dependence on $\eta$. However, rather than use
(\ref{YY}) to obtain a limit, it is preferable to use 
the likelihood method.

 Just as \he4 and \li7 were sufficient to
determine a value for $\eta$,  a limit on $N_\nu$ can be obtained
as well \cite{fo,oth2,oth3,sark2}. The likelihood approach
utilized above can be extended to include $N_\nu$ as a free parameter.
Since the light element abundances can be computed as functions
of both $\eta$ and $N_\nu$,  the
likelihood function can be defined by \cite{oth2}
replacing the quantity $Y_{\rm BBN}\left(\eta\right)$ in eq. (\ref{gau})
with $Y_{\rm BBN}\left(\eta,N_\nu\right)$ to obtain ${L^{^4{\rm He}}}_{\rm
total}(\eta,N_\nu)$.
Again, similar expressions are needed for \li7 and D. 

\begin{figure}[htbp]
\hspace{0.5truecm}
\includegraphics[width=15pc]{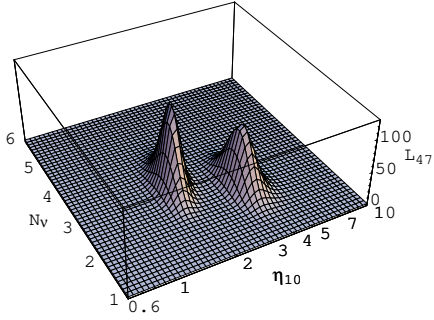}
\caption{$L_{47}(N_\nu, \eta)$ for observed abundances given by
                 eqs. (\protect\ref{he4} and \protect\ref{li}).}
\label{fig:oth31}
\end{figure}

\begin{figure}[htbp]
\hspace{0.5truecm}
\includegraphics[width=15pc]{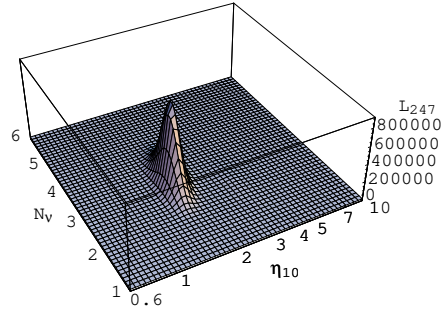}
\caption{$L_{247}(N_\nu, \eta)$ for observed abundances including high
D/H.}
\label{fig:oth32}
\end{figure}

\begin{figure}[htbp]
\hspace{0.5truecm}
\includegraphics[width=15pc]{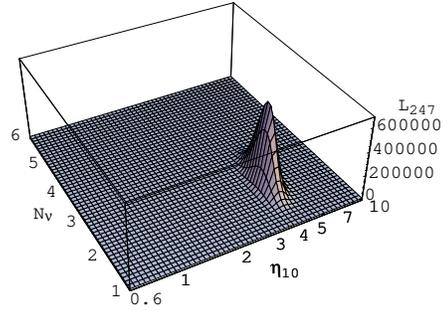}
\caption{$L_{247}(N_\nu, \eta)$ for observed abundances including low D/H.}
\label{fig:oth33}
\end{figure}

A three-dimensional view of the combined likelihood functions \cite{oth3} 
is shown in Figure \ref{fig:oth31}.  In this case the high and low
$\eta$ maxima of Figure \ref{fig:fig2}, show up as peaks in the
$L-\eta-N_\nu$ space.  The likelihood function is labeled $L_{47}$ (and
$L_{247}$ when D/H is included). In Figures \ref{fig:oth32} and
\ref{fig:oth33} the corresponding likelihood functions
$L_{247}$ with high and low D/H are shown. Once again one sees an effect
of including D/H is to eliminate one of the \li7 peaks.  
Furthermore, unlike the case discussed in section 6, the likelihood
distribution for low D/H is just as strong as that for high D/H, albeit
at a lower value of $N_\nu$.

\begin{figure}[htbp]
\hspace{0.5truecm}
\includegraphics[width=15pc]{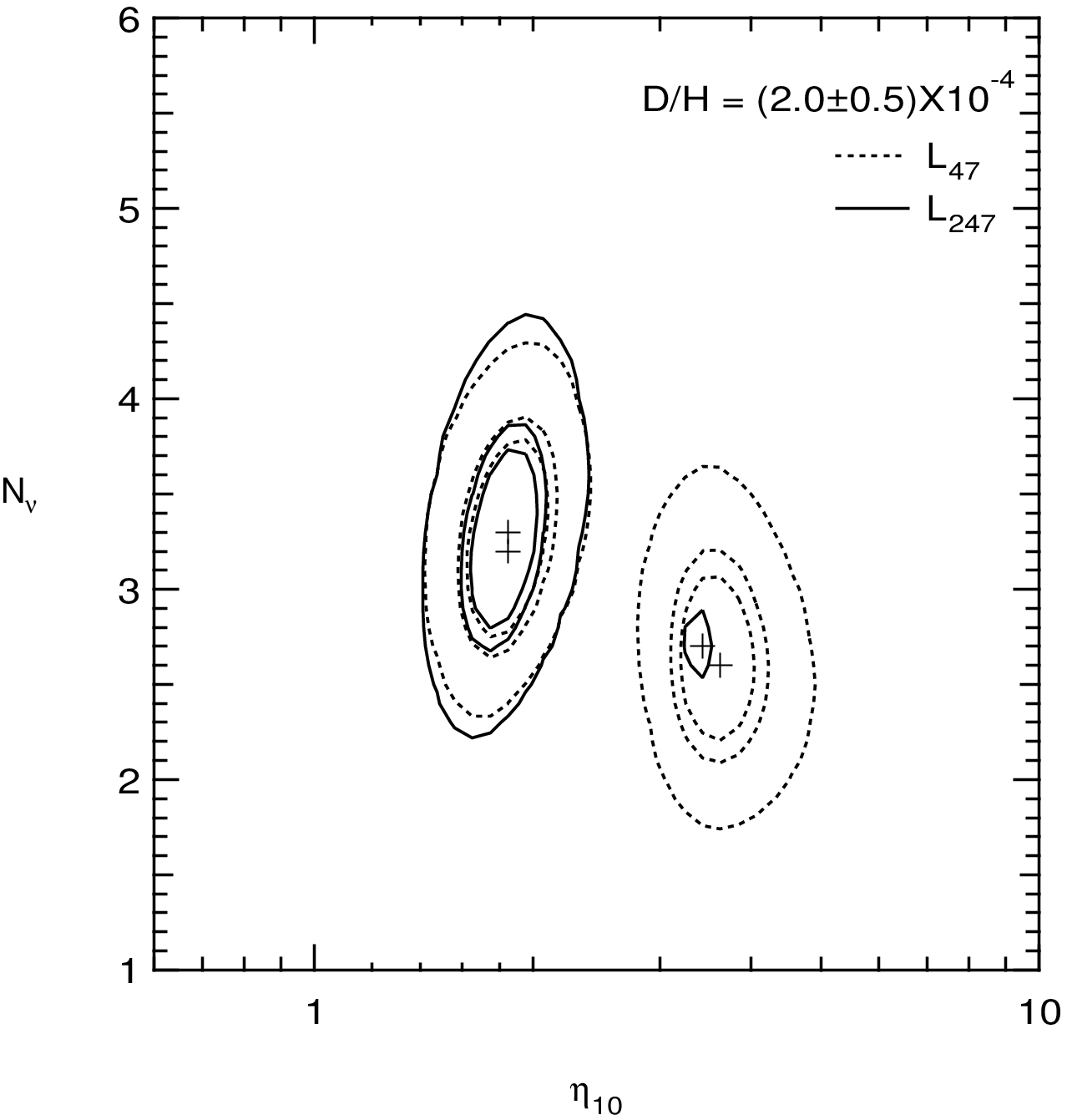}
\caption{50\%, 68\% \& 95\% C.L. contours of $L_{47}$ and
                 $L_{247}$ where observed abundances are given by
                 eqs. (\protect\ref{he4} and  \protect\ref{li}), and
high D/H.}
\label{fig:oth34}
\end{figure}

\begin{figure}[htbp]
\hspace{0.5truecm}
\includegraphics[width=15pc]{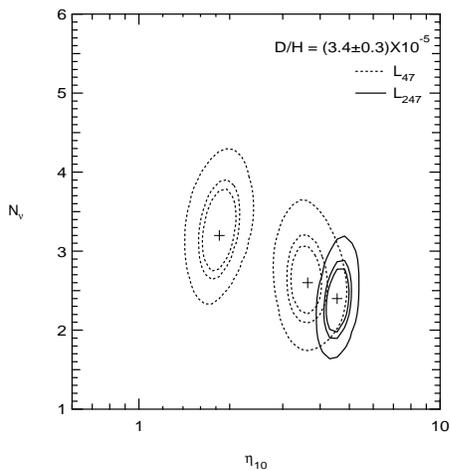}
\caption{50\%, 68\% \& 95\% C.L. contours of $L_{47}$ and
                 $L_{247}$ where observed abundances are given by
                  eqs. (\protect\ref{he4} and  \protect\ref{li}), and
low D/H.}
\label{fig:oth35}
\end{figure}

 The peaks of the distribution as
well as the allowed ranges of $\eta$ and $N_\nu$ are  
more easily discerned in the 
contour plots of Figures \ref{fig:oth34} and \ref{fig:oth35} which show
the 50\%, 68\% and 95\% confidence level contours in $L_{47}$ and
$L_{247}$ for high and low D/H as indicated.   The crosses show the
location of the  peaks of the likelihood functions.
$L_{47}$ peaks at $N_\nu=3.2$, $\eta_{10}=1.85$  and at $N_\nu=2.6$,
$\eta_{10}=3.6$.  The 95\% confidence level allows the following ranges
in $\eta$ and $N_\nu$
\beq
1.7\le N_\nu\le4.3  \qquad \qquad
1.4\le\eta_{10}\le 4.9 
\eeq
Note however that the ranges in $\eta$ and $N_\nu$ are strongly
correlated as is evident in Figure \ref{fig:oth34}.

With high D/H, $L_{247}$ 
peaks at $N_\nu=3.3$, and also at $\eta_{10}=1.85$. 
In this case
the 95\% contour gives the ranges
\beq
2.2\le N_\nu\le4.4 \qquad \qquad
1.4\le\eta_{10}\le 2.4
\eeq
Note that within the 95\% CL range, there is also a small area 
with $\eta_{10} = 3.2 - 3.5$ and $N_\nu = 2.5-2.9$.

Similarly, for  low D/H, $L_{247}$ 
peaks at $N_\nu=2.4$, and $\eta_{10}=4.55$. 
The 95\% CL upper limit is now $N_\nu < 3.2$, and the range for 
$\eta$ is $ 3.9 < \eta_{10} < 5.4$.  It is important to stress that with
the increase in the determined value of D/H \cite{bty} in the low D/H
systems, these abundances are now consistent with the standard model
value of $N_\nu = 3$ at the 2 $\sigma$ level.


\section{Summary} 

To summarize, 
I would assert that one
can conclude that the present data on the abundances of the light element
isotopes are consistent with the standard model of big bang 
nucleosynthesis. Using
the isotopes with the best data, \he4 and
\li7, it is possible to constrain the theory and obtain a best set of
values for the baryon-to-photon ratio of $\eta_{10}$ and the
corresponding  value for $\Omega_B h^2$ 
\beq
\begin{array}{ccccc}
1.55 & < & \eta_{10} & < &  4.45 \qquad 95\% {\rm CL} \nonumber \\
.006 & < & \Omega_B h^2 & < & .016  \qquad 95\% {\rm CL}
\label{res2}
\end{array}
\eeq
For $0.4 < h < 1$, we have a range $ .006 < \Omega_B < .10$.
This is a rather low value for the baryon density
 and would suggest that much of the galactic dark matter is
non-baryonic \cite{vc}. These predictions are in addition 
consistent with recent
observations of D/H in quasar absorption systems which show a high value.
Difficulty remains however, in matching the solar \he3 abundance, suggesting a
problem with our current understanding of galactic chemical evolution or the
stellar evolution of low mass stars as they pertain to \he3.

\vskip .2in

\noindent This work was supported in part by
DoE grant DE-FG02-94ER-40823 at the University of Minnesota.

\end{document}